\documentclass[12pt]{article}

\usepackage{amsthm,amssymb,amsmath,mathrsfs,stmaryrd}

\title{\Large Singletons and their maximal symmetry algebras}

\author{Xavier Bekaert
\\ 
{\small Laboratoire de Math\'ematiques et Physique Th\'eorique}\\
{\small Unit\'e Mixte de Recherche $6083$ du CNRS}\\
{\small F\'ed\'eration de Recherche $2964$ Denis Poisson}\\
{\small Universit\'e Fran\c{c}ois Rabelais, Parc de Grandmount}\\
{\small 37200 Tours, France}\\
{\tt\small xavier.bekaert@lmpt.univ-tours.fr}
}

\date{}

\begin{document}

\maketitle

\abstract{
Singletons are those unitary irreducible modules of the
Poincar\'e or (anti) de Sitter group that can be lifted to unitary modules of the conformal group. Higher-spin algebras
are the corresponding realizations of the universal enveloping algebra of the
conformal algebra on these modules. These objects
appear in a wide variety of areas of theoretical physics: AdS/CFT correspondence, electric-magnetic duality, higher-spin
multiplets, infinite-component Majorana equations, higher-derivative symmetries, etc. Singletons and higher-spin algebras are reviewed through a list of their many equivalent definitions in order to approach them from various perspectives. The focus of this introduction is on the symmetries of a singleton: its maximal algebra and the manifest realization thereof.



\vspace{8mm}\noindent\textit{\footnotesize 
Synthesis of various talks and lectures 
presented at the ``6th Alsacian meeting in mathematics and physics''
(Strasbourg, France; November 2009), ``Seminars on higher spins at Mons'' (Mons, Belgique; March 2010), ``7th spring school and workshop on quantum field theory \& Hamiltonian systems'' (Craiova \& Calimanesti, Romania; May 2010), ``International conference on non-commutative structures and non-relativistic (super) symmetries''
(Tours, France; June 2010), conference ``Quantum field theory and gravity'' (Tomsk, Russia; July 2010), ``6th mathematical physics meeting: summer school and conference on modern mathematical physics'' (Belgrade, Serbia; September 2010).}
}

\thispagestyle{empty}

\clearpage 

\section{Plan of a singleton sightseeing tour}

The celebrated singletons are rather ``remarkable representations'', as coined by Dirac in his seminal paper \cite{Dirac:1963ta} on the subject. Indeed, these representations of the anti de Sitter spacetime isometry group possess several surprising properties which are so exceptional that they distinguish singletons from all other such representations. Several of these properties are reviewed here, thereby providing an elementary introduction to singletons through a list, presumably inexhaustive, of their distinct but equivalent definitions. Exhibiting the many faces of singletons could give some flavor of their ubiquitous appearances in such seemingly unrelated areas of mathematical physics as the AdS/CFT correspondence, the hydrogen atom spectrum, the electric-magnetic duality, the infinite-component Majorana equation, etc. An exhaustive bibliographical survey of the wide range of results and applications for singletons is by no means attempted here.\footnote{The bibliography has been deliberately focused either on some recent general reviews with indications of the precise location of the relevant information, or on some old seminal papers, in order to give some flavor of the early history, though from a modern viewpoint. I do apologize to the experts for the incompleteness of the bibliography.} On the contrary the main focus of this short introduction is on the symmetry algebras of bosonic singletons in any dimension and on their manifest realizations. No prior knowledge of singletons is assumed, but some familiarity with the representation theory of Lie algebras is welcome. The plan is as follows:

In order to be as self-contained as possible, the isometry groups of the anti de Sitter spacetime and its conformal boundary are quickly reviewed in Section 2, as well as the corresponding representation theory classifying the elementary particles that may live on these spaces. Then comes the section 3 which presents many faces of singletons: lowest weight modules (subsection 3.1), multiplicity free modules (subsect 3.2), irreducible modules of isometry subalgebras (subsect 3.3), fields on the conformal boundary (subsect 3.4), fields on the ambient space (subsect 3.5) and kernels of the Howe dual algebra (subsect 3.6).
The simplest example of singleton is the scalar one and it will serve throughout this review as a useful illustration. 
Finally, the section 4 reviews the various definitions of bosonic higher-spin algebras: as realizations of universal enveloping algebras (subsect 4.1),
as centralizers of Howe dual algebras (subsect 4.2), as invariants of Howe dual algebras (subsect 4.3), as algebras of symmetries
(subsect 4.4). The final message of this tour is that although higher-spin algebras can be defined in many mathematically equivalent ways, their most physical interpretation is presumably as maximal symmetry algebras of free singletons (as motivated by the AdS/CFT correspondence in the exotic strongly-curved/weakly-coupled regime).
 
\section{Elementary particles\\ on anti de Sitter spacetime}\label{elementary}

\subsection{Anti de Sitter spacetime}

The most transparent realization of $AdS_{n+1}$ ($n\geqslant 1$) is via a global isometric embedding in a flat ambient space:
\begin{itemize}
	\item The \textbf{ambient space} ${\mathbb R}^{n,2}$ is endowed with the
	\begin{itemize}
		\item Cartesian coordinates $X^A$ (where $A=0,0^\prime,1,2,\ldots,n$) and
		\item (``mostly plus'') metric $\eta_{AB}\,=\,\,$diag$(-1,-1,+1,+1,\ldots,+1)$.
	\end{itemize}

	\item The \textbf{anti de Sitter spacetime} $AdS_{n+1}$ is the codimension one quadric (more precisely, the one-sheeted hyperboloid)
	$$\eta_{AB}X^AX^B\,=\,-R^2\,,$$
	where $R>0$ is the curvature radius, endowed with the induced metric.
\end{itemize}

\noindent So the \textbf{isometry algebra} is manifestly the real 
Lie algebra $$\mathfrak{o}(n,2)=\mbox{span}_{\mathbb R}
\{\textsc{J}_{AB}\}$$ which can be {presented}
\begin{itemize}
	\item by its {generators} (the ambient ``angular momenta'') $$\textsc{J}_{AB}=-\textsc{J}_{BA}\quad  \mbox{(where $A,B=0,0^\prime,1,2,\ldots,n$).}$$
	\item modulo the {commutation relations}
$$\left[\textsc{J}_{AB},\textsc{J}_{CD}\right]\,=\,i\,\eta_{BC}\textsc{J}_{AD}\, +\, \mbox{antisymmetrizations}\,.$$
\end{itemize}
and is \textbf{linearly realized} on ${\mathbb R}^{n,2}$ through the generators (the ambient ``orbital angular momenta'') $$\textsc{J}_{AB}=\textsc{X}_{A}\textsc{P}_B-\textsc{X}_{B}\textsc{P}_A$$
where $$\textsc{P}_A\,=\,-\,i\,\frac{\partial}{\partial X^A}.$$

Usually, one of the timelike direction of ${\mathbb R}^{n,2}$, say $0^\prime$, is particularized.
Equivalently, one of the points of $AdS_{n+1}$, say of coordinates $X^{0^\prime}=R$ and $X^a=0$ (where $a=0,1,2,\ldots,n$), is particularized.
Then the generators decompose in two sets:
\begin{itemize}
\item The stabilizer of $X^a=0$, i.e. the \textbf{Lorentz subalgebra} $$\mathfrak{o}(n,1)=\mbox{span}_{\mathbb C}\{\textsc{J}_{ab}\}$$ which can be presented
	\begin{itemize}
		\item by its generators $\textsc{J}_{ab}=-\textsc{J}_{ba}$ (where $a,b=0,1,2,\ldots,n$)
		\item modulo the commutation relations
		$$\left[\textsc{J}_{ab},\textsc{J}_{cd}\right]\,=\,i\,\eta_{bc}\textsc{J}_{ad}\, +\, \mbox{antisymmetrizations}\,.$$
	\end{itemize}

\item The \textbf{transvections} (the displacements) generated by $\Gamma_a\,:=\,R\,\textsc{J}_{0^\prime a}$ and satisfying
the commutation relations
		$$\left[\Gamma_a,\Gamma_b\right]\,=\,\frac{i}{R^2}\,\textsc{J}_{ab}\,.$$
\end{itemize}
The transvections generators transform in the vector representation of the Lorentz subalgebra $\mathfrak{o}(n,1)$, as can be seen in the commutation relations
$$\left[\textsc{J}_{ab},\Gamma_c\right]\,=\,i\,\eta_{bc}\Gamma_a\, +\, \mbox{antisymmetrization}\,.$$
At the level of isometry algebras, the flat spacetime limit  $AdS_{n+1}\stackrel{R\to \infty}{\longrightarrow}{\mathbb R}^{n,1}$ translates into the
In\"onu-Wigner contraction $\mathfrak{o}(n,2)\stackrel{R\to \infty}{\longrightarrow}\mathfrak{io}(n,1)$ where the transvections become the translations of the Poincar\'e algebra $\mathfrak{io}(n,1)={\mathbb R}^{n}
\niplus \mathfrak{o}(n-1,1)$.

\subsection{Conformal boundary}

\subsubsection{Conformal isometries of Minkowski spacetime}

A \textbf{conformal metric} is an equivalence class of a metric under the equivalence relation
$$g^\prime_{\mu\nu}(x)\sim \Omega(x)g_{\mu\nu}(x)\qquad \mbox{(where $\mu,\nu=0,1,2,\ldots,n-1$)}$$
with $\Omega(x)>0$ for all $x^\mu$ while
a \textbf{conformal isometry} is a diffeomorphism such that the metric transforms as $$g^\prime_{\mu\nu}(x')\,=\,\Omega(x)\,g_{\mu\nu}(x)\,. $$
In other words, a (conformal) isometry is a diffeomorphism that preserves the (conformal) metric. In particular, conformal isometries preserve the causal structure of spacetime (i.e. the gender of tangent vectors).

From now on, one will restrict to the case $n\geqslant 3$. In such case, the finite conformal isometries of the \textbf{Minkowski spacetime} ${\mathbb R}^{n-1,1}$ are generated (see e.g. \cite{DiFrancesco:1997nk} for a proof) by the:
\begin{itemize}
	\item \textbf{Lorentz transformations} $x^{\prime\mu}=\Lambda^\mu{}_\nu x^\nu$, where $\Lambda\in O(n-1,1)$\,,
	\item \textbf{Translations}: $x^{\prime\mu}=x^\mu+a^\mu$, where $a\in {\mathbb R}^{n-1,1}$\,,
	\item \textbf{Dilatations}: $x^{\prime\mu}=\lambda\, x^\mu$, where $\lambda\in{\mathbb R}_0:={\mathbb R}-\{0\}$\,, 
\end{itemize}
together with, either the:
\begin{itemize}	
	\item \textbf{Special conformal transformations}: $x^{\prime\mu}= \frac{x^\mu+ x^2b^\mu}{1+2b_\mu x^\mu+b^2x^2}$, 
	
	where $b\in {\mathbb R}^{n-1,1}$,
\end{itemize}
or the:
\begin{itemize}	
	\item \textbf{Inversion}: $x^{\prime\mu}= \frac{x^\mu}{x^2}$\,.
\end{itemize}
The special conformal transformations are the most difficult to visualize but they may be understood indirectly from the following property: \textit{Special conformal transformations are conjugate to translations, via the inversion.} 

The algebra of the infinitesimal conformal isometries of ${\mathbb R}^{n-1,1}$ is $\mathfrak{o}(n,2)$ but this is far from obvious in terms of the Cartesian coordinates $x^\mu$.
Moreover, the last two transformations (special conformal transformations and inversions) are not well defined everywhere on ${\mathbb R}^{n-1,1}$ because they map some points ``at infinity'' (when the denominator vanish). In order to make the conformal isometries well defined globally, it is necessary to complete the Minkowski spacetime ${\mathbb R}^{n-1,1}$ by adding ``points at infinity''. The corresponding conformal compactification of ${\mathbb R}^{n-1,1}$ can be identified with the conformal boundary of $AdS_{n+1}$.

\subsubsection{Conformal boundary of anti de Sitter spacetime}

The most transparent realization of the conformal boundary $\partial AdS_{n+1}$ of the anti de Sitter spacetime $AdS_{n+1}$ is via its global (conformal isometric) embedding in the projectivization of the ambient space ${\mathbb R}^{n,2}$:

\begin{itemize}
	\item The {ambient space} is now the projective space ${\mathbb P}({\mathbb R}^{n,2})\cong {\mathbb R}{\mathbb P}^{n+1}$ endowed with the
	\begin{itemize}
		\item \textbf{Homogeneous coordinates} $X^A$ (where $A=0,0^\prime,1,2,\ldots,n$),
		\item Equivalence relation $X^A\sim \lambda \,X^A$ (for any $\lambda\in{\mathbb R}_0$)
		\item Conformal metric (i.e. the equivalence class of) $\eta_{AB}$
	\end{itemize}
	As usual, points of the projective space ${\mathbb P}({\mathbb R}^{n,2})$ are rays of ${\mathbb R}^{n,2}$.

	\item The \textbf{Dirac hypercone}\footnote{This construction is the Euclidean analogue of the ``M\"obius model'' in the mathematical literature. It was introduced a long time ago in physics by Dirac in a paper \cite{Dirac:1936fq} which still remains a splendid introduction to the ambient formulation. A beautiful result of modern conformal geometry is the generalization of the ambient formulation to curved conformal spaces by Fefferman and Graham \cite{Fefferman}.} is the embedded codimension one quadric (more precisely, the null cone with the tip excluded)
	$$\eta_{AB}X^AX^B\,=\,0$$
	quotiented by the equivalence relation, endowed with the induced conformal metric. This conformal space is the conformal boundary of the anti de Sitter spacetime $AdS_{n+1}$. 
	\end{itemize}
Geometrically, the points of $\partial AdS_{n+1}$ are null rays of the ambient space ${\mathbb R}^{n,2}$.	Topologically, the space $\partial AdS_{n+1}$ is, essentially, homeomorphic to $S^1\times S^{n-1}$ (where $S^p$ denotes a hypersphere of dimension $p$).
	Heuristically, the boundary of the anti de Sitter spacetime is the asymptotic (i.e. located ``at infinity'') region of intersection between the hyperboloid and the hypercone.

The \textbf{conformal isometry algebra} of the conformal boundary $\partial AdS_{n+1}$ is manifestly the real Lie algebra $\mathfrak{o}(n,2)$ linearly realized on ${\mathbb P}({\mathbb R}^{n,2})$ through the (``ambient orbital angular momentum'') generators. More concretely, it is {linearly realized} on ${\mathbb R}^{n,2}$ through the generators $\textsc{J}_{AB}=\textsc{X}_{[A}\textsc{P}_{B]}$, where the square bracket denotes the antisymmetrization.

\subsubsection{Conformal isometries revisited}

The \textbf{light-cone coordinates} $$X^\pm:=X^{0^\prime}\pm X^{n}$$
together with the \textbf{inhomogeneous coordinates}
$$x^\mu:=X^\mu/X^-\qquad \mbox{(where $\mu=0,1,2,\ldots,n-1$)}$$
provide a convenient parametrization of the Dirac hypercone
in a neighborhood such that $X^-\neq 0$. Indeed, the intersection of the hypercone $-X^+X^-+\eta_{\mu\nu}X^\mu X^\nu=0$ with a hyperplane $X^-=\,$constant$\,\neq 0$ is the codimension one paraboloid of equation $X^+=X^-\,x^2$ parametrized by the inhomogeneous coordinates.
The hyperplane
$X^-=0$ may be taken as the ``hyperplane at infinity'' to be added to the affine space ${\mathbb R}^{n+1}$ in order to construct the (projective) ambient space ${\mathbb R}{\mathbb P}^{n+1}$.
If one identifies the conformal boundary of $AdS_{n+1}$ with the conformal compactification of ${\mathbb R}^{n-1,1}$ then the ``hyperplane at infinity'' is indeed particularized.

The conformal isometries decompose as follows (see e.g. \cite{Siegel} for a short review):
\begin{itemize}
	\item The ambient isometries $X^{\prime A}=\Lambda^A{}_{B}X^B$ preserving the hyperplane $X^-=\,$constant$\,\neq 0$, i.e. the
	\begin{itemize}
		\item {Lorentz transformations}:
		$$X^{\prime +}=X^+,\,\, X^{\prime \mu}=\Lambda^\mu{}_\nu X^\nu \quad\Longleftrightarrow\quad x^{\prime\mu}=\Lambda^\mu{}_\nu x^\nu$$
		where $\Lambda\in O(n-1,1)$.
		\item {Translations}:
		$$X^{\prime +}=X^++2\,a_\mu X^\mu+a^2\,X^-,\,\, X^{\prime \mu}=X^\mu + a^\mu X^- \quad\Longleftrightarrow\quad x^{\prime\mu}=x^\mu+a^\mu$$
\end{itemize}
	\item The ambient isometries in the plane $0^\prime n\leftrightarrow +-$ which preserve the hyperplane at infinity $X^-=0$, i.e. the  {dilatations}:
		$$X^{\prime +}=\lambda\,X^+,\,\,X^{\prime -}=\lambda^{-1}\,X^-,\,\, X^{\prime \mu}= X^\mu\quad\Longleftrightarrow\quad x^{\prime\mu}=\lambda x^\mu$$
	\item The remaining transformations, i.e. the {special conformal transformations}:
	$$X^{\prime +}=X^+,\,X^{\prime -}=X^-+2\,b_\mu X^\mu+b^2\,X^+,\, X^{\prime \mu}=X^\mu + b^\mu X^+$$
	$$\qquad \Longleftrightarrow\quad x^{\prime\mu}= \frac{x^\mu+ x^2b^\mu}{1+2b_\mu x^\mu+b^2x^2}$$
	\item The reflection through the hyperplane $X^n=0$, i.e. the {inversion}:
	$$X^{\prime +}=X^-,\,\,X^{\prime -}= X^+,\,\, X^{\prime \mu}=X^\mu \quad\Longleftrightarrow\quad x^{\prime\mu}= \frac{x^\mu}{x^2}\,.$$
\end{itemize}
So the infinitesimal generators decompose as follows:
\begin{itemize}
	\item The stabilizer of any hyperplane $X^-=\,$constant$\,\neq 0$, i.e. the \textbf{Poincar\'e subalgebra}
	$$\mathfrak{io}(n-1,1)=\mbox{span}
	_{\mathbb R}
	\{\textsc{P}_\mu,\textsc{J}_{\mu\nu}\}={\mathbb R}^{n}
\niplus \mathfrak{o}(n-1,1)$$ which can be presented
	\begin{itemize}
		\item by its generators $$\textsc{P}_\mu:=\textsc{J}_{+\mu}/2\,,\quad \textsc{J}_{\mu\nu}=-\textsc{J}_{\nu\mu} \quad(\mu,\nu=0,1,2,\ldots,n-1)$$
		\item modulo the commutation relations
		$$\left[\textsc{P}_{\mu},\textsc{P}_{\nu}\right]\,=\,0\,,$$
		$$\left[\textsc{P}_{\mu},\textsc{J}_{\nu\rho}\right]\,=\,i\,\eta_{\mu\nu}\textsc{P}_{\rho} +\, \mbox{antisymmetrization}\,,$$
		$$\left[\textsc{J}_{\mu\nu},\textsc{J}_{\rho\sigma}\right]\,=\,i\,\eta_{\nu\rho}\textsc{J}_{\mu\sigma}\, +\, \mbox{antisymmetrizations}\,.$$
	\end{itemize}
	\item The generator of ambient isometries in the plane $0^\prime n\leftrightarrow +-$ which preserve the hyperplane at infinity, i.e. the generator of {dilatation} $$\Delta:= \textsc{J}_{+-}$$

	\item The remaining generators, corresponding to the infinitesimal {special conformal transformations}
	$$\textsc{K}_\mu:=\textsc{J}_{-\mu}$$
\end{itemize}

\subsubsection{Distinct constant curvature spacetimes\\ as an identical conformal space}

Actually, the conformal boundary of $AdS_{n+1}$ may be identified with any of the three constant curvature spacetimes (supplemented by ``points at infinity''). These three spacetimes are geometrically realized as quadrics obtained by intersecting the hypercone with an affine hyperplane:
\begin{itemize}
	\item \textbf{Minkowski spacetime} ${\mathbb R}^{n-1,1}$ endowed with the Cartesian coordinates $x^\mu$ as before. 
	\begin{itemize}
		\item \textbf{Paraboloid}: intersection with a hyperplane orthogonal to a light-like direction, say $X^-=\,$constant$\,\neq 0$
		\item Isometry algebra: Poincar\'e algebra $\bf\mathfrak{io}(n-1,1)$
	\end{itemize}

	\item \textbf{de Sitter spacetime} $dS_n$
	\begin{itemize}
		\item \textbf{Hyperboloid}: intersection with a hyperplane orthogonal to a time-like direction, say $X^{0^\prime}=R\neq 0$
		\item Isometry algebra: $\bf\mathfrak{o}(n,1)$
	\end{itemize}

	\item \textbf{Anti de Sitter spacetime} $AdS_n$
	\begin{itemize}
		\item \textbf{Hyperboloid}: intersection with a hyperplane orthogonal to a space-like direction, say $X^n=R\neq 0$
		\item {Isometry algebra}: $\bf\mathfrak{o}(n-1,2)$
	\end{itemize}
\end{itemize}
The conformal compactifications of the three distinct constant curvature spacetimes ${\mathbb R}^{n-1,1}$, $dS_n$ and $AdS_n$ are identical: they reproduce the flat conformal space $\partial AdS_{n+1}$. Indeed, all of these spacetimes are conformally flat and they possess the same {conformal isometry algebra} $\mathfrak{o}(n,2)$.
It is important to emphasize this point because, although most of the time the conformal boundary $\partial AdS_{n+1}$ is identified with the conformal compactification of Minkowski spacetime ${\mathbb R}^{n-1,1}$, from the point of view of conformal geometry it can equivalently be taken to be the conformal compactification of (anti) de Sitter spacetime $(A)dS_n$. This remark is useful because most results which will be mentioned here equally apply to all constantly curved spacetimes.

\subsection{Unitary irreducible representations}

\subsubsection{Elementary particles as irreducible modules}

Let $\cal M$ be a maximally symmetric spacetime.
A celebrated insight due to Wigner (see e.g. the section 2 of \cite{Bekaert:2006py} for a review) states that there exists a one-to-one correspondence between the set of (equivalence classes of)
\begin{itemize}
	\item[(i)] unitary representations of the isometry algebra of $\cal M$ on a module (``representation space'') $\cal H$ and
	\item[(ii)] linear relativistic wave equations describing the free propagation of a quantum particle on $\cal M$.
\end{itemize}
The unitary module $\cal H$ is the Hilbert space of physical states (i.e. of inequivalent solutions of the wave equation). 
This identification allows to define a \textbf{particle} as a unitary module of the spacetime isometry group.
Moreover, any unitary module which is reducible is actually completely reducible. Thence, there is no loss of generality in restricting the attention to unitary irreducible modules.
So one defines an \textbf{elementary particle} as an irreducible unitary module of the isometry group of the spacetime $\cal M$.

So the classification of the free elementary particles in the anti de Sitter spacetime ${\cal M}=AdS_{n+1}$
tantamounts to the classification of the irreducible unitary modules of $\mathfrak{o}(n,2)$. Strictly speaking, the Hilbert space of physical states is usually the direct sum of two irreducible modules: the ones with either positive or negative energy corresponding respectively to the particle and its antiparticle. Therefore, if the sign of the energy is not specified in the sequel, then the direct sum of the positive energy module and its conjugate should be understood.

\subsubsection{Isometry algebra of anti de Sitter spacetime}

The \textbf{maximal compact subalgebra} $\mathfrak{o}(2) \oplus \mathfrak{o}(n)$ of the real Lie algebra $\mathfrak{o}(n,2)$ corresponds to the
\begin{itemize}
	\item \textbf{time translations} generated by the (conformal) \textbf{Hamiltonian} $\textsc{E}=\textsc{J}_{0' 0}$
	\item \textbf{spatial rotations} generated by $\textsc{J}_{ij}$ (where $i,j=1,2,\ldots,n$)
\end{itemize}
acting in a natural way on the boundary $\partial AdS_{n+1}$ homeomorphic to $S^1\times S^{n-1}$.
The remaining generators can be recast in the form of \textbf{ladder operators} $$\textsc{J}^{\pm}_j=\textsc{J}_{0j}\mp i \textsc{J}_{0'j}\,,$$ raising or lowering the \textbf{energy} (= eigenvalue of $\textsc{E}$) by one unit. Indeed, the real Lie algebra $\mathfrak{o}(n,2)$ can be presented equivalently
\begin{itemize}
	\item by the generators $\textsc{E}$, $\textsc{J}_i^{\pm}$, $\textsc{J}_{jk}$ (where $i,j,k=1,2,\ldots,n$)
	\item modulo the {commutation relations}
\begin{eqnarray}
&&\left[\textsc{E}, \textsc{J}_i^{\pm}\right]=\pm  \textsc{J}_i^{\pm} ~~~;~~~ \left[\textsc{J}_{ij},\textsc{J}^{\pm}_k\right]= 2i\delta_{k[j} \textsc{J}^{\pm}_{i]}\nonumber\\
&&\left[\textsc{J}^-_i, \textsc{J}^+_j\right]=2(i\textsc{J}_{ij}+\delta_{ij}\textsc{E})\nonumber\\
&&\left[\textsc{J}_{ij},\textsc{J}_{kl}\right]=i\delta_{jk}\textsc{J}_{il} \,+\, \mbox{antisymmetrizations}\nonumber
\end{eqnarray}		
\end{itemize}

\subsubsection{Orthogonal algebra}

The classification of the irreducible unitary modules of $\mathfrak{o}(n,2)$ requires the knowledge of the classification of the irreducible unitary modules of $\mathfrak{o}(n)$, which can be summarized as follows (see e.g. the section 3 of \cite{Bekaert:2006py} for more details and references):

\vspace{2mm}\noindent\textbf{Irreducible unitary modules of the orthogonal group}:
\textit{
Let $n\geqslant 3$ be a positive integer and $[\frac{n}2]$ denote the integer part of $\frac{n}2$.
Any unitary irreducible module ${\cal D}_{\ell_1,\ldots,\ell_{[\frac{n}2]}}$ of $O(n)$ is a finite-dimensional highest (and lowest) weight module which is
\begin{itemize}
	\item either tensorial or spinorial,
	\item labeled by a partition of an integer $$|\,\ell|=\ell_1+\ell_2+\ldots+\ell_{[\frac{n}2]}$$ in $[\frac{n}2]$ parts (where $\ell_1\geqslant\ell_2\geqslant\ldots\geqslant\ell_{[\frac{n}2]}\geqslant 0$).
\end{itemize}
The converse is also true.}
\vspace{1mm}

A partition of $|\,\ell|$ in $p$ parts is usually depicted as a \textbf{Young diagram} made of $|\,\ell|$ boxes arranged in $p$ left-justified rows of non-increasing lengths
$$\ell_1\geqslant\ell_2\geqslant\ldots\geqslant\ell_p\geqslant 0\,.$$
A renowned example of $\mathfrak{o}(n)$-module is the \textbf{spin-$s$ module} ${\cal D}_s$ corresponding to a partition of $|\,\ell|=s\in\mathbb N$ in one part. It corresponds either to a
\begin{itemize}
	\item tensorial module ${\cal D}_s$ of $\mathfrak{o}(n)$ spanned by the components of a symmetric traceless tensor of rank $s$, or a
	\item spinorial module ${\cal D}_{s+1/2}$ of $\mathfrak{o}(n)$ spanned by the components of a symmetric (gamma)-traceless tensor-spinor of rank $s$.
\end{itemize}

\subsubsection{Irreducible modules of the isometry algebra}

A \textbf{Verma module} ${\cal V}(E_0;\ell_1,\ldots,\ell_{[\frac{n}2]})$ of $\mathfrak{o}(n,2)$ for $E_0$ positive (or negative)
\begin{itemize}
	\item is obtained by the action of the \textbf{universal enveloping algebra} ${\cal U}\big(\mathfrak{o}(n,2)\big)$ on the
	\item \textbf{Lowest (or highest) weight vector} $\vert E_0;\ell_1,\ldots,\ell_{[\frac{n}2]}\rangle$ of $\mathfrak{o}(n,2)$, which is defined as a
	\begin{itemize}
		\item Lowest energy $E_0$ state
		$$\textsc{J}_i^-\vert E_0;\ell_1,\ldots,\ell_{[\frac{n}2]}\rangle\,=\,0$$
		(or highest energy state $\textsc{J}_i^+\vert E_0;\ell_1,\ldots,\ell_{[\frac{n}2]}\rangle\,=\,0$)	
		\item Lowest (and highest) weight vector of $\mathfrak{o}(n)$ labeled by 
		$$\ell_1\geqslant\ell_2\geqslant\ldots\geqslant\ell_{[\frac{n}2]}\geqslant 0.$$
	\end{itemize}
\end{itemize}
For the general definitions and properties of Verma modules, the reader may look for instance at the concise review in \cite{Fuchs}.
The ground states of energy $E_0$ span an irreducible finite-dimensional $\mathfrak{o}(n)$-module labelled by the above partition.

For physical reasons, an elementary particle on anti de Sitter spacetime is taken to be a positive-energy (lowest weight) unitary module, while its antiparticle is its opposite counterpart, so a negative-energy (highest weight) unitary module. Both cases can be described as \textbf{extremal weight} unitary modules.

\vspace{2mm}\noindent\textbf{Irreducible unitary modules of the isometry algebra}:
\textit{Any extremal weight irreducible module ${\cal D}(E_0;\ell_1,\ldots,\ell_{[\frac{n}2]})$ of $\mathfrak{o}(n,2)$
is a quotient of the Verma module ${\cal V}(E_0;\ell_1,\ldots,\ell_{[\frac{n}2]})$ by its maximal submodule ${\cal V}(E^\prime_0;\ell^\prime_1,\ldots,\ell^\prime_{[\frac{n}2]})$.
}

\noindent Unitarity imposes some restrictions (see e.g. \cite{Metsaev:1997nj} and refs therein) on the possible values of the extremal energy when
the ground state $\mathfrak{o}(n)$-module carries the
\begin{itemize}
	\item trivial representation ($\ell_1=\ell_2=\ldots=\ell_{[\frac{n}2]}=0$): either
\begin{itemize}
	\item[$*$] $E_0=0$, corresponding to the trivial $\mathfrak{o}(n,2)$-module, or
	\item[$*$] $|E_0|\geqslant \frac{n}2-1$ corresponding to scalar field $\mathfrak{o}(n,2)$-modules,
\end{itemize}
	\item spinor representation ($\ell_1=\ell_2=\ldots=\ell_{[\frac{n}2]}=0$): 
\begin{itemize}
	\item[$*$] $|E_0|\geqslant \frac{n-1}2$ corresponding to the spinor field $\mathfrak{o}(n,2)$-modules,
\end{itemize}
	\item generalized spin-$s$ representation, labeled by a Young diagram with a upper rectangle made of $k$ rows with length $[s]$ (i.e. a partition such that $[s]=\ell_1=\ell_2=\ldots=\ell_k>\ell_{k+1}$ with $1\leqslant k\leqslant [n/2]$):  
\begin{itemize}
	\item[$*$] $|E_0|\geqslant s+n-k-1$ corresponding to tensorial (or spinorial) modules according to whether $2s$ is even (or odd).
\end{itemize}
\end{itemize}
The difference $\tau:=|E_0|-s$ is the \textbf{twist} of such extremal weight irreducible $\mathfrak{o}(n,2)$-modules. The former unitarity bounds imply that $\tau\geqslant \frac{n}2-1$.

For more details on the $n=3$ unitary irreducible modules, an excellent pedagogical introduction is \cite{deWit:1999ui}.
For more details on the generic construction of $\mathfrak{o}(n,2)$ unitary irreducible modules, one may look at the appendices in \cite{Iazeolla:2008ix}.

\section{Singletons: various definitions}

\subsection{Singletons as lowest weight modules}

Singletons form the very exceptional subclass of the irreducible unitary modules of the algebra $\mathfrak{o}(n,2)$ that saturate the previous unitary bound on the twist: $\tau=\frac{n}2-1= 0$. Equivalently, they saturate the unitarity bound on the extremal energy $E_0$
and their ground states are characterized by a rectangular Young diagram made of $[s]$ columns of height $k=\frac{n}2$, where $s$ may differ from $0$ or $1/2$ if and only if $n$ is even.

The group-theoretic definition of singletons dates back to the seminal works of Dirac \cite{Dirac:1963ta} and Flato \& Fr\o nsdal \cite{Flato:1978qz}. The higher-dimensional generalization\footnote{\label{osc}Related to the isomorphism $\mathfrak{o}(3,2)\cong \mathfrak{sp}(4)$, the $n=3$ singletons admit a realization in terms of bosonic oscillators
\cite{Dirac:1963ta} which is called the ``metaplectic'' representation of the symplectic group by mathematicians. 
``Singletons'' are sometimes defined as the natural generalization of this construction, \textit{c.f.} the nice review \cite{Gunaydin}. This other possibility is extremely interesting in its own but this definition is, in general, distinct from the definition that we follow here for generic $n$.} was further developped by a variety of authors (a self-contained and rather complete treatment of the generic case can be found in \cite{Angelopoulos:1997ij}):

\vspace{2mm}
\noindent\textbf{Definition:}\textit{
A positive-energy \textbf{singleton of $\bf AdS_{n+1}$} is a lowest weight unitary irreducible module of $\mathfrak{o}(n,2)$ such that:
\begin{itemize}
	\item When $n$ is odd, the spin is
	\begin{itemize}
		\item either $s=\frac12$: ${\cal D}(\frac{n-1}2;\frac12)$ called \textbf{Di} or \textbf{spinor singleton}
		\item or $s=0$: ${\cal D}(\frac{n}2-1;0)$ called \textbf{Rac} or \textbf{scalar singleton}
	\end{itemize}
	\item When $n$ is even, the (generalized) spin $s$ is 
	\begin{itemize}
		\item any (half)integer:
	${\cal D}(s+\frac{n}2-1\,;\,[s],\ldots,[s])$ called \textbf{spin $s$ singleton} and labeled by a partition in $\frac{n}2$ equal parts, i.e. a rectangular Young diagram made of $\frac{n}2$ rows of length $[s]$.
	\end{itemize}
\end{itemize}
}
\noindent\vspace{2mm}From now on, whenever singletons of spin $s\geqslant 1$ will be mentioned, the integer $n$ will always be implicitly assumed to be even, as it should.

\vspace{2mm}
As a nice illustration, the scalar singleton may deserve a more detailed discussion. A short review of scalar singletons can be found in \cite{Engquist:2007vj}.
The {scalar singleton} corresponds to the case of a lowest weight vector $\vert E_0;0\rangle$ of $\mathfrak{o}(n,2)$
annihilated by all generators of the $\mathfrak{o}(n)$ subalgebra:
$$
(\textsc{E}-E_0)\vert E_0;0\rangle =0~~~,~~~ \textsc{J}_{ij}\vert E_0;0\rangle=0~~~,~~~ \textsc{J}^-_i\vert E_0;0\rangle=0\,.
$$
Thus the Verma module is
$$
{\cal V}(E_0;0)=\mbox{span}_{\mathbb R}\left\{\textsc{J}^+_{i_1} \dots \textsc{J}^+_{i_s}\vert E_0;0\rangle\,\mid s\in{\mathbb N}\right\}
$$
It can be shown (see e.g. \cite{deWit:1999ui} for $n=3$) that unitarity implies $E_0\geqslant \frac{n}2-1$ (or $E_0=0$ which corresponds to the trivial representation of $\mathfrak{o}(n,2)$\,).
For the special value $E_0=\frac{n}2-1$ saturating the unitarity bound,
the vector $\delta^{ij}\textsc{J}^+_i \textsc{J}^+_j \vert E_0;0\rangle$ is a primitive null vector. The scalar singleton is the unitary module obtained by quotienting the maximal submodule
$$
{\cal V}\Big(\frac{n}2+1;0\Big)\cong \mbox{span}_{\mathbb R}\left\{\delta^{i_1i_2}
\textsc{J}^+_{i_1}\dots \textsc{J}^+_{i_s}\vert E_0;0\rangle\,\mid s\in{\mathbb N}\right\}\subset {\cal V}\Big(\frac{n}2-1;0\Big)
$$
from the Verma module. Concretely, this corresponds to factoring out the trace terms from the Verma module:
\begin{eqnarray}
&&\mathcal D\Big(\frac{n}2-1;0\Big)=\mathcal V\Big(\frac{n}2-1;0\Big)\bigg/\mathcal V\Big(\frac{n}2+1;0\Big)\nonumber\\
&&\cong\mbox{span}_{\mathbb R}\left\{\, \textsc{J}^+_{i_1}\dots \textsc{J}^+_{i_s}\vert\, n/2-1;0\rangle\,\,\mid\,\, \delta^{i_1i_2}
\textsc{J}^+_{i_1}\dots \textsc{J}^+_{i_s}\vert\, n/2-1;0\rangle\sim 0\,\right\}\nonumber
\end{eqnarray}

\subsection{Singletons as multiplicity free modules}

A nice corollary of the previous description is that under the restriction of $\mathfrak{o}(n,2)$ to its subalgebra $\mathfrak{o}(n)$, the scalar singleton module is decomposable as follows
$$
\mathcal D\Big(\frac{n}2-1;0\Big)\Big|_{\mathfrak{o}(n)}\cong \bigoplus\limits_{s\in\mathbb N} {\cal D}_s
$$
where each irreducible $\mathfrak{o}(n)$-module ${\cal D}_s$ is an eigenspace of distinct energy
$$
\Big(\textsc{E}-\big(s+\frac{n}2-1\big)\Big){\cal D}_s=0
$$
and appears with multiplicity one.
Actually, the terminology ``singleton'' originates\footnote{The author thank V.K. Dobrev for pointing out to him that the name ``singleton'' comes from this fact.} from this absence of degeneracy of $\mathfrak{o}(n)$-modules in the spectrum of the $n=3$ ``Di'' and ``Rac'' modules, as observed initially in \cite{Ehrman}.
This property generalizes to any dimension \cite{Angelopoulos:1999bz}.

\vspace{2mm}\noindent\textbf{Theorem (Angelopoulos \& Laoues, 2000)}:
\textit{
A positive energy {singleton} of $AdS_{n+1}$ is a non-trivial lowest weight unitary irreducible module of $\mathfrak{o}(n,2)$ such that 
its reduction to the compact orthogonal subalgebra $\mathfrak{o}(n)$ is multiplicity free. The corresponding weights of the maximal compact subalgebra $\mathfrak{o}(2)\oplus\mathfrak{o}(n)$ lie along a line in the weight diagram. More precisely,
$$
\mathcal D\Big(s+\frac{n}2-1\,;\,[s],[s],\ldots,[s]\Big)\Big|_{\mathfrak{o}(n)}\cong \bigoplus\limits_{t\in\mathbb N} {\cal D}_{[s]+t,[s],\ldots,[s]}
$$
where each irreducible $\mathfrak{o}(n)$-module ${\cal D}_{[s]+t,[s],\ldots,[s]}$ is an eigenspace of distinct energy
$$
\left(\textsc{E}-\big(\,[s]+t+\frac{n}2-1\big)\right){\cal D}_{[s]+t,[s],\ldots,[s]}=0
$$
and appears with multiplicity one.
Moreover, singletons are those non-trivial extremal weight unitary irreducible modules of $\mathfrak{o}(n,2)$ such that their reduction to a compact Cartan subalgebra is multiplicity free: there is at most a single linearly independent eigenvector (i.e. a ``singlet'') for each weight.}
\vspace{1mm}

As a side remark, one may observe that the previous properties are the roots of the earliest appearances, but in disguised form, of the $n=3$ singleton module in particle physics through the so-called infinite-component equations and the spectrum-generating algebras. These ideas were pushed forward in the sixties by the ``dynamical group'' research programme which revived the seminal work of Majorana \cite{Majorana}. A very concise account of Majorana's publication itself and of the history of infinite-component wave equations can be found in \cite{Casalbuoni:2006fa}.
Such constructions were recently revisited in \cite{Bekaert:2009pt}
from a modern and shifted perspective (including a discussion of the higher-dimensional cases $n\geqslant 4$) which will briefly reviewed now.

Because of the commutation relations of the algebra $\mathfrak{o}(n,2)$
$$\left[\textsc{J}_{ab},\textsc{J}_{cd}\right]\,=\,i\,\eta_{bc}\textsc{J}_{ad}\, +\, \mbox{antisymmetrizations}$$
$$\left[\Gamma_a,\Gamma_b\right]\,=\,i\,\textsc{J}_{ab}$$
(where one has put $R=1$ since the curvature radius has no physical interpetation in the present context), the generators $\Gamma_a$ can be reinterpreted as infinite-dimensional Dirac matrices from which the ``spinning'' generators of $\mathfrak{o}(n-1,1)$ are constructed as: $\textsc{J}_{ab}\,=\,-i\,\Gamma_{[a}\Gamma_{b]}$.
If the wave function $\psi(x)$ on the Minkowski spacetime ${\mathbb R}^{n,1}$ takes values in the scalar singleton module $\mathcal D(n/2-1;0)$,
then the Dirac-like (i.e. the Majorana infinite-component) equation
$$(\Gamma^a \textsc{P}_a \,-\,M)\psi(x)=0$$
possesses a discrete spectrum of massive solutions. As one can see by looking at this equation in the rest frame such that $(\textsc{P}_a \,-\,m\,\delta_a^0)\psi(x)=0$, their masses $m$ are related to their spin $s$ by the relation
$$m\,=\,\frac{M}{s+\frac{n}2-1}$$
since the eigenvalue of $\Gamma_0=\textsc{E}$ on the spin-$s$ $\mathfrak{o}(n)$-submodule ${\cal D}_s$ is equal to $s+\frac{n}2-1$.
Unfortunately, this ``Regge trajectory'' is decreasing so the infinite-component Majorana equation is physically unsatisfactory. Furthermore, tachyonic and continuous spin particles also appear in the spectrum.

As a curiosity, one may mention the following observation \cite{Bekaert:2009pt} based on the above-mentioned decomposition of the scalar singleton module: Consider an infinitely degenerate spectrum of massive particles on the flat spacetime ${\mathbb R}^{n,1}$ with equal
mass for all spins (multiplicity one), i.e. a horizontal Regge trajectory.
In the rest frame, the massive particle of spin $s$ is described by the irreducible $\mathfrak{o}(n)$-module ${\cal D}_s$, therefore the
infinite tower of particles fits in an irreducible multiplet of $\mathfrak{o}(n,2)$ corresponding to the scalar singleton module
$\mathcal D(\frac{n}2-1;0)$.
One might speculate that such an infinite multiplet could come from a highly degenerate (and exotic) spontaneous symmetry breaking of the anti de Sitter isometry algebra \cite{Bekaert:2009pt}.

\subsection{Singletons as irreducible modules\\ of isometry subalgebras}

The following theorem of Angelopoulos \& Laoues \cite{Angelopoulos:1997ij} extends to any $n\geqslant 3$ the previous result for the case $n=3$ of 
Angelopoulos, Flato, Fr\o nsdal \& Sternheimer \cite{Angelopoulos:1980wg}.

\vspace{2mm}\noindent\textbf{Theorem (Angelopoulos \& Laoues, 1998)}:
\textit{A positive (or negative) energy {singleton} of $AdS_{n+1}$ is a non-trivial lowest (or highest) weight unitary irreducible $\mathfrak{o}(n,2)$-module that remains irreducible (or, at most, splits in two irreducible modules) under restriction to any of the following subalgebras: $\mathfrak{io}(n-1,1)$, $\mathfrak{o}(n,1)$ and $\mathfrak{o}(n-1,2)$.
Conversely, a {singleton} on $\partial AdS_{n+1}$ is a unitary irreducible module (at most, a sum of two such modules) of any of the previous subalgebras, that can be lifted to a unitary module of $\mathfrak{o}(n,2)$.
}

\vspace{2mm}
Heuristically, this means that \textbf{singletons of} $\bf AdS_{n+1}$ are those fields:
\begin{itemize}
	\item whose local physical degrees of freedom sit on its conformal boundary, so one may also call them \textbf{singletons on} $\bf\partial AdS_{n+1}$, and
	\item which are preserved by the conformal symmetries in spacetime dimension $n$.
\end{itemize}
The absence of local degrees of freedom for singletons of $AdS_{n+1}$ is confirmed by the fact that, upon the In\"onu-Wigner contraction $\mathfrak{o}(n,2)\rightarrow\mathfrak{io}(n,1)$ corresponding to the flat spacetime limit $AdS_{n+1}\rightarrow{\mathbb R}^{n,1}$, they carry a trivial representation of the translation subalgebra ${\mathbb R}^{n+1}$ in the Poincar\'e algebra \cite{Flato:1978qz}.

Consequently, the extremal weight unitary irreducible $\mathfrak{o}(n,2)$-modules are of two types: the ones that are
\begin{itemize}
	\item {singletons on} $\partial AdS_{n+1}$, describing \textbf{elementary particles of a $\bf CFT_n$} (= $n$-dimensional conformal field theory), or
  \item \textit{not} {singletons on} $\partial AdS_{n+1}$, describing \textbf{elementary particles in the interior of $\bf AdS_{n+1}$}.
\end{itemize}
The fact that elementary particles may live either on the boundary or in the interior of anti de Sitter spacetime is the very basis of the
$AdS_{n+1}/CFT_n$ correspondence.

\vspace{2mm}Let the conformal boundary $\partial AdS_{n+1}$ be identified with the conformal compactification of ${\mathbb R}^{n-1,1}$.
The following theorem was found by Siegel in \cite{Siegel:1988gd} but the entirely complete and rigorous proof was given later in \cite{Angelopoulos:1997ij}.

\vspace{2mm}\noindent\textbf{Theorem (Siegel, 1989)}:
\textit{A positive-energy {singleton} on $\partial AdS_{n+1}$ is a positive-energy massless unitary irreducible module of $\mathfrak{io}(n-1,1)$ induced (\`a la Wigner) by a finite dimensional irreducible representation of the stabilizer $\mathfrak{io}(n-2)$ labeled by a partition in $\frac{n}2-1$ equal parts, i.e. by a rectangular Young diagram made of $\frac{n}2-1$ rows of length $[s]$.
}

\noindent\vspace{1mm}
When seen as a representation of the Poincar\'e subalgebra $\mathfrak{io}(n-1,1)$, the spin $s$ singleton on the conformal compactification of ${\mathbb R}^{n-1,1}$ is called, for
\begin{itemize}
	\item $n=4$, the \textbf{helicity $s$ representation}, and for
	\item higher even $n$ and $s\geqslant 1$, a \textbf{spin $s$ duality-symmetric representation} when $n/2$ is even or \textbf{chiral representation} when $n/2$ is odd,	because the corresponding fieldstrength span an irreducible $\mathfrak{o}(n-1,1)$-module described by a rectangular Young diagram made of $\frac{n}2$ rows for which Hodge self-duality may be defined (more information on this point is provided in the next subsection).
\end{itemize}

\vspace{1mm}
Now let the conformal boundary $\partial AdS_{n+1}$ be identified with the conformal compactification of $AdS_n$.
The analogue of the previous theorem was obtained by Metsaev in \cite{Metsaev:1995jp} (see also \cite{Angelopoulos:1997ij} for a proof).

\vspace{2mm}\noindent\textbf{Theorem (Metsaev, 1995)}:
\textit{A positive-energy singleton on $\partial AdS_{n+1}$ is a positive-energy unitary irreducible $\mathfrak{o}(n-1,2)$-module ${\cal D}(s+\frac{n}2-1;[s],\ldots,[s])$
saturating the unitarity bound of $\mathfrak{o}(n-1,2)$-modules when $s\neq 0$ (and of $\mathfrak{o}(n,2)$-modules when $s=0$) and whose lowest energy $\mathfrak{o}(n-1)$-module is labeled by a partition in $\frac{n}2-1$ equal parts, i.e. by a rectangular Young diagram made of $\frac{n}2-1$ rows of length $[s]$.
}

\vspace{2mm}\noindent\textbf{Remark:} It is important to stress that a singleton \textit{on} $\partial AdS_{n+1}$ is \textit{not} a singleton \textit{of} $AdS_n$, but of $AdS_{n+1}$. Indeed, a singleton on $\partial AdS_{n+1}$ has lowest energy $s-1+n/2$ while a singleton of $AdS_n$ has lowest energy $s-1+(n-1)/2$. Moreover, spin $s\geqslant 1$ singletons on $\partial AdS_{n+1}$ exist only for $n$ even while spin $s\geqslant 1$ singletons of $AdS_n$ exist only for $n$ odd.
\vspace{1mm}

A similar theorem holds for the conformal compactification of $dS_n$ as well \cite{Angelopoulos:1997ij}.

\subsection{Singletons as fields on\\ the conformal boundary}

Singletons live on the conformal boundary so they can be described as fields on the corresponding compactified spacetimes.
The simplest example is the {scalar singleton} which can be described as a massless (i.e. harmonic) scalar field $\phi(x^\mu)$ on Minkowski spacetime ${\mathbb R}^{n-1,1}$ of conformal weight $1-\frac{n}2$ so that the d'Alembert equation
$$\square_{{\mathbb R}^{n-1,1}}\, \phi(x)=0\quad\mbox{is preserved by the conformal algebra}\,\,\, \mathfrak{o}(n,2)\,.$$
Equivalently, the scalar singleton may be described on $(A)dS_n$ through a linear wave equation involving the conformal (or Yamabe) Laplacian $$\Big(\square_{(A)dS_n}\pm \frac{n(n-2)}{4\,R^2}\Big) \phi(x)=0$$
where $\square$ denotes the Laplace-Beltrami operator and the $\pm$ symbol refer to the sign of the spacetime curvature.

The spin-$s$ singleton $\mathfrak{io}(n-1,1)$-module can be realized as a space of {harmonic irreducible multiforms} \cite{Bandos:2005mb} (see e.g. the section 5 of \cite{Bekaert:2006py} for a review of the general construction of Poincar\'e modules). For $n=4$, this construction reproduces the famous Bargmann-Wigner equations which are known to be conformally symmetric since a long time ago \cite{Gross:1964}. 
For definiteness, one will focus on tensorial singletons, i.e. integer spin $s\in\mathbb N$. Let $\theta_i^\mu$ be a set (where $\mu,\nu=0,1,2,\ldots,n-1$ and $i=1,2,\ldots, s-1,s$) of fermionic coordinates
$$\theta_i^\mu\theta_j^\nu+\theta_j^\nu\theta_i^\nu=0 \,,$$
on $\Pi({\mathbb R}^{n-1,1}\otimes {\mathbb R}^s)$ (where $\Pi$ reverses the Grassmann parity).

\vspace{2mm}
\noindent\textbf{Definitions:} \textit{A (differential) \textbf{multiform} on Minkowski spacetime ${\mathbb R}^{n-1,1}$ is a function $\psi(x^\mu,\theta^\nu_i)$ on the superspace
${\mathbb R}^{n-1,1}\,\oplus\,\Pi({\mathbb R}^{n-1,1}\otimes{\mathbb R}^s)$, i.e. tensor fields on ${\mathbb R}^{n-1,1}$ with components
described by a product of $s$ columns. 
Moreover, a multiform is:
\begin{itemize}
	\item \textbf{closed} if it is annihilated by all operators $d_i=\vartheta^\mu_i\frac{\partial}{\partial x^\mu}$
	\item \textbf{coclosed} if it is annihilated by all operators $d^\dagger_i=\frac{\partial}{\partial \vartheta_\mu^i}\frac{\partial}{\partial x^\mu}$.
	\item \textbf{harmonic} if it is closed and coclosed.
\end{itemize}
}

\vspace{1mm}\noindent The differential multiforms for $s\geqslant 2$ generalize the usual differential forms ($s=1$). The components of a multiform span an irreducible $GL(n)$-module described by a rectangular Young diagram made of $s$ columns
and $\frac{n}2$ rows iff it is annihilated by the operators
$\theta^\mu_i \frac{\partial}{\partial \theta^\mu_j}\,-\,\delta_i^j\,\frac{n}{2}$ that span the algebra $\mathfrak{gl}(s)$.
Moreover, it is further irreducible under $O(n-1,1)$ iff it is also annihilated by the operators
$\theta^\mu_i \theta^j_\mu$ and $\frac{\partial}{\partial \theta^\mu_i}\frac{\partial}{\partial \theta_\mu^j}$
which, together with the previous ones, span the algebra $\mathfrak{o}(2s)$.

\vspace{2mm}\noindent\textbf{Poincar\'e covariant equations for singletons \cite{Bandos:2005mb}:}
\textit{The spin-$s$ singleton $\mathfrak{io}(n-1,1)$-module can be realized as a space of multiforms $\psi(x,\theta)$ on the Minkoswki spacetime ${\mathbb R}^{n-1,1}$ 
which are harmonic and whose components span an irreducible $\mathfrak{o}(n-1,1)$-module ${\cal D}_{[s],\ldots,[s]}$ labeled by a rectangular Young diagram made of $[s]$ columns and $\frac{n}2$ rows.
}

\noindent\vspace{1mm}
\textbf{Remark:} The multiform associated with a {singleton} of spin $s\geqslant 1$ is physically interpreted as its \textbf{fieldstrength} (or \textbf{curvature tensor}). An irreducible $O(n-1,1)$-module labeled by a rectangular Young diagram made of $[s]$ columns and $\frac{n}2$ rows decomposes a sum of two irreducible $\mathfrak{o}(n-1,1)$-modules when $n/2$ is odd. This subtlety is related to the involutive property of the Hodge operator. In order to treat both cases uniformly, one should consider the complexification of the $O(n-1,1)$-module when $n/2$ is even. Then both modules are eigenspaces (of eigenvalue $\pm 1$) of the involutive Hodge duality, with a factor $i$ included when $n/2$ is even (see e.g. the article \cite{Deser:1997mz} containing a concise introduction to the $s=1$ case). 

\vspace{1mm}
So the singleton modules of spin $s\geqslant 1$ are either said to be {duality-symmetric} when $n/2$ is even or chiral when $n/2$ is odd.
This duality properties extend to constantly curved spacetimes, therefore the singletons of spin $s\geqslant 1$ are those finite-component unitary irreducible representations of one of the isometry algebras $\mathfrak{io}(n-1,1)$, $\mathfrak{o}(n,1)$ or $\mathfrak{o}(n-1,2)$ which are either duality-symmetric or chiral.
This deep connection between conformal symmetry and electric-magnetic duality somehow explains the appearance of singletons in many celebrated models of high-energy theoretical physics, such as maximally supersymmetric theories.

\subsection{Singletons as fields on the ambient space}

The main drawback of the description of singletons as fields on the conformal boundary (presented in the previous subsection) is that the conformal symmetry is not manifest (dilatation symmetry is obvious but not the special conformal and inversion symmetries).
To circumvent this defect, one may describe singletons as fields on the ambient space. 
Such a description was initiated by Dirac in \cite{Dirac:1936fq} and can be summarized for the scalar singleton as follows (see e.g. the section 3 of \cite{Eastwood:2002su} for a review of this elegant construction):

On the one hand, any space of functions of the inhomogeneous coordinates on ${\mathbb R}{\mathbb P}^{n+1}$ can be realized in terms of the homogeneous coordinates as a space of homogeneous functions on ${\mathbb R}^{n+2}$ of some fixed degree. On the other hand, any space of functions on the null cone can be realized as a space of equivalence classes of functions on the ambient space modulo the functions which vanish on the null cone. The homogeneity degree is fixed by the requirement that the Laplace-Beltrami operator on the ambient space ${\mathbb R}^{n,2}$ preserves the latter equivalence relation, so that this operator induces the conformal Laplacian on the conformal boundary $\partial AdS_{n+1}$.

\vspace{2mm}\noindent\textbf{Ambient construction of the scalar singleton (Dirac, 1936)}:
\textit{The {scalar singleton} $\mathfrak{o}(n,2)$-module can be realized as a space of functions $\Phi(X)$ on the ambient space ${\mathbb R}^{n,2}$ which are
\begin{itemize}
	\item harmonic: $\square_{{\mathbb R}^{n,2}}\, \Phi(X)=0$
	\item of homogeneity degree $1-\frac{n}2$: $(X^A\partial_A+\frac{n}2-1)\Phi(X)=0$
	\item quotiented by the equivalence relation $$\Phi(X)\sim \Phi(X)\,+\,(X^AX_A)\,\Xi(X)$$ where $\Xi(X)$ is of homogeneity degree $-1-\frac{n}2$.
\end{itemize}
}

\vspace{1mm}
\noindent\textbf{Remark:} The operators $\square\,$, $X\cdot\partial_X+\frac{n+2}2\,$, $X^2$ are called (first class) \textbf{constraints} and they span the symplectic algebra $\mathfrak{sp}(2)$. This property will be made manifest via Howe duality. These first-class constraints find a natural interpretation in the  ``two-time physics'' research programme of Bars (see e.g. \cite{Bars:2001xv}). Actually, all constraints can equivalently be imposed on the physical states.
\vspace{1mm}

\vspace{2mm}\noindent\textbf{Ambient construction of the scalar singleton}:
\textit{The {scalar singleton} module can be realized as a space of distributions $$\Psi(X):=\delta(X^2)\Phi(X)$$ on ambient space ${\mathbb R}^{n,2}$ which are
\begin{itemize}
	\item harmonic: $\square\, \Psi(X)=0$
	\item of homogeneity degree $-1-\frac{n}2$: $(X\cdot\partial+\frac{n+2}2)\Psi(X)=0$
	\item annihilated by the quadratic form: $\,X^2\, \Psi(X)=0$
\end{itemize}
}

\vspace{1mm}The generalization of this construction to any spin \cite{Arvidsson:2006fq} can be performed in the language of multiform.
Let $\vartheta_i^A$ be a set (where $i=1,2,\ldots, s-1,s$) of fermionic coordinates
$$\vartheta_i^A\vartheta_j^B+\vartheta_j^B\vartheta_i^A=0 \,,$$
on $\Pi({\mathbb R}^{n,2}\otimes {\mathbb R}^s)$.

\vspace{2mm}
\noindent\textbf{Definitions:}\textit{An \textbf{ambient multiform} is
\begin{itemize}
	\item a multiform on the ambient space ${\mathbb R}^{n,2}$, i.e. a function $\Psi(X^A,\vartheta^B_i)$ on the superspace
${\mathbb R}^{n,2}\,\oplus\,\Pi({\mathbb R}^{n,2}\otimes{\mathbb R}^s)$.
	\item \textbf{tangent to the anti de Sitter spacetime $AdS_{n+1}$} if it is annihilated by all operators 	$X^A\frac{\partial}{\partial \vartheta^A_i}$.
	\item \textbf{tangent to the conformal boundary $(\partial AdS)_n$} if it is annihilated by all operators
$X^A \frac{\partial}{\partial \vartheta^A_i}$ and	$\frac{\partial}{\partial X^A} \frac{\partial}{\partial \vartheta^i_A}$.
\end{itemize}
}

\noindent\vspace{1mm}The definitions of (co)closure and harmonicity for ambient multiforms are the analogues of the ones for spacetime multiforms.

\vspace{2mm}
\noindent\textbf{Ambient construction (Arvidsson \& Marnelius, 2006):}
The {tensorial singleton} $\mathfrak{o}(n,2)$-module can be realized as a space of multiforms $\Psi(X,\vartheta)$ on the ambient space ${\mathbb R}^{n,2}$
\begin{itemize}
	\item which are
\begin{itemize}
	\item harmonic
	\item of homogeneity degree $-1-\frac{n}2$
	\item annihilated by $X^2$
	\item tangent to the conformal boundary
\end{itemize}
	\item whose components span an irreducible $\mathfrak{o}(n,2)$-module described by a rectangular Young diagram made of $s$ columns and $\frac{n}2+1$ rows.
\end{itemize}

\vspace{1mm}This formulation is appealing because conformal symmetry is manifest, unfortunately the price to pay is that locality is not manifest any more. However, there exists a formulation \cite{Bekaert:2009fg} where both conformal invariance and locality are
manifest. This is made possible by an ambient space construction in the fiber
rather than in the spacetime, along the lines of the parent approach \cite{Barnich:2004cr}.
The BRST construction of \cite{Bekaert:2009fg} appears closely related to the tractor formalism \cite{tractor} in conformal geometry used in the description of fields on curved ambient space (see \textit{e.g.} \cite{Gover:2009vc} for some applications).

\vspace{1mm}
\noindent\textbf{Remark}: The various operators $\square\,$, $X\cdot\partial_X+\frac{n+2}2\,$, $X^2$, $\frac{\partial}{\partial X}\cdot \vartheta_i$,
 $\frac{\partial}{\partial X}\cdot \frac{\partial}{\partial \vartheta_i}$,
 $X\cdot \frac{\partial}{\partial \vartheta_i}$,
 $\frac{\partial}{\partial X}\cdot \frac{\partial}{\partial \vartheta_i}$.
 $\vartheta_i\cdot \frac{\partial}{\partial \vartheta_j}\,-\,\delta_i^j\,\frac{n+2}{2}$, $\vartheta_i\cdot \vartheta_j$, and  $\frac{\partial}{\partial \vartheta_i}\cdot \frac{\partial}{\partial \vartheta_j}$ which annihilate the module span the
orthosymplectic superalgebra $\mathfrak{osp}(2s|2)$ of constraints. This superalgebra finds a natural interpretation, on the mathematical side, in terms of Howe duality, and, on the physical side, in terms of the $\mathfrak{o}(2s)$ extended supersymmetric spinning particle (see e.g. \cite{Arvidsson:2006fq,Bastianelli:2008nm} and refs therein).

\subsection{Singletons as kernels of\\ the Howe dual algebra}

An important message is that the orthosymplectic $\mathfrak{osp}(2s|2)$ (super)algebra of constraints annihilating the spin-$s$ singleton module is the Howe dual of the conformal algebra $\mathfrak{o}(n,2)$ acting on the singleton irreducible module. For a concrete description of Howe duality, one may look e.g. at the the section 3 of the review \cite{Bekaert:2005vh}. For the sake of simplicity, let us turn back to the scalar singleton.

Let $T^*{\mathbb R}^{n,2}$ be the (trivial) \textbf{cotangent bundle} of the ambient space with canonical
\begin{itemize}
	\item \textbf{Darboux coordinates} $Y^A_\alpha=(X^A,P_A)$ (where $\alpha=1,2$)
	\item \textbf{Poisson bracket}
$$\{Y_\alpha^A,Y^B_\beta\}=\varepsilon_{\alpha\beta}\,\eta^{AB}\quad\Longleftrightarrow\quad \{X^A,P_B\}=\delta^A_B$$
\end{itemize}
where $\varepsilon_{\alpha\beta}$ is the symplectic form of $\mathfrak{sp}(2)$.
The \textbf{Weyl algebra} $A_{n+2}$ is the algebra of (polynomial) differential operators $\textsc{O}(\textsc{X},\textsc{P})$ on the ambient space ${\mathbb R}^{n,2}$, where $\textsc{P}_A\,=\,-\,i\,\partial/\partial X^A$. The Weyl algebra is isomorphic to the space of \textbf{Weyl symbols} $O(X,P)$, i.e. (polynomial) functions on the cotangent bundle $T^*{\mathbb R}^{n,2}$, endowed with the \textbf{Moyal star product} $*=\exp i\{\,\,,\,\}$. 

On the one hand, the algebra $\mathfrak{o}(n,2)=\mbox{span}
\{\textsc{L}^{AB}\}$ is linearly realized on ${\mathbb R}^{n,2}$ via the generators
$\textsc{L}^{AB}=\textsc{X}^{A}\textsc{P}^B-\textsc{X}^{B}\textsc{P}^A$ whose Weyl symbols are the bilinears
$$L^{AB}=\varepsilon^{\alpha\beta}Y_\alpha^A Y^B_\beta=X^AP^B-X^BP^A$$
On the other hand, the 
Lie algebra $$\mathfrak{sp}(2)=\mbox{span}
\{\textsc{u}_{\alpha\beta}\}$$ can be presented
\begin{itemize}
	\item by its {generators} $\textsc{u}_{\alpha\beta}=\textsc{u}_{\beta\alpha}$ (where $\alpha,\beta=1,2$)
	\item modulo the {commutation relations}
$$\left[\textsc{u}_{\alpha\beta},\textsc{u}_{\gamma\delta}\right]\,=\,i\,\varepsilon_{\beta\gamma}\textsc{u}_{\alpha\delta}\, +\, \mbox{symmetrizations}\,.$$
\end{itemize}
The Weyl symbols of the operators $\square\,$, $X\cdot\partial_X+\frac{n+2}2\,$, $X^2$ are the bilinears
$$U_{\alpha\beta}=\eta_{AB}Y_\alpha^A Y^B_\beta$$
In both cases, the Moyal commutators (or Poisson brackets) of generators reproduce the corresponding commutation relations.
These respective realizations of $\mathfrak{o}(n,2)$ and $\mathfrak{sp}(2)$ are maximal commutants in the algebra of quadratic Weyl symbols: they form a  \textbf{Howe dual pair}.

\vspace{2mm}\noindent\textbf{Ambient construction of the scalar singleton}:
\textit{The {scalar singleton} $\mathfrak{o}(n,2)$-module is a space of distributions $\Psi(X)$ on the ambient space ${\mathbb R}^{n,2}$ which are annihilated by the $\mathfrak{sp}(2)$ algebra Howe dual to $\mathfrak{o}(n,2)$ in the algebra of linear operators on ${\mathbb R}^{n,2}$:
$$\textsc{u}_{\alpha\beta}\Psi(X)=0$$
}

The generalization to any integer spin $s\in\mathbb N$ is analogous \cite{Bekaert:2009fg}: 
The Grassmann even indices $A,B$ will still correspond to the $(n+2)$-dimensional ambient space ${\mathbb R}^{n,2}$ with metric $\eta^{AB}$ but the letters $\alpha,\beta$ will now be superindices corresponding to a $(2|2s)$-dimensional symplectic superspace
$$T^*{\mathbb R}^{1|s}\cong {\mathbb R}^{2|2s}$$ with symplectic form ${\cal J}_{\alpha\beta}$.
The symplectic form on the superspace ${\mathbb R}^{2|2s}$
can be seen as a metric form on the superspace ${\mathbb R}^{2s|2}\cong \Pi ({\mathbb R}^{2|2s})$
with opposite Grassmann parity. Therefore, the symplectic form ${\cal J}_{\alpha \beta}$ is manifestly preserved by the orthosymplectic algebra
$\mathfrak{osp}\,(2s\,|\,2)$.
The multiforms are functions on the
superspace $${\mathbb R}^{n,2}\,\oplus\,\Pi({\mathbb R}^{n,2}\otimes{\mathbb R}^s)\cong {\mathbb R}^{n+2\,|\,s(n+2)}$$ with
\begin{itemize}
	\item $n+2$ even coordinates $X^A$ on ${\mathbb R}^{n,2}$
  \item $s(n+2)$ odd coordinates $\vartheta_i^A$ on $\Pi({\mathbb R}^{n,2}\otimes{\mathbb R}^s)$.
\end{itemize}
Let $(P_A|\pi^i_B)$ be the conjugates of the supercoordinates $(X^A|\theta^B_i)$.
The phase (super)space coordinates on the cotangent bundle $T^*{\mathbb R}^{n+2|s(n+2)}$ are collectively denoted by $$Z^A_{\alpha}:=(X^A,P_B|\,\theta_i^A,\pi^j_B)$$ where
the superindex $\alpha$ takes $2+2s$ values.
The graded Poisson bracket originating from the symplectic
structure on the phase superspace is
$$\{{Z_\alpha^A},{Z^B_\beta}\}=\eta^{AB}{\cal J}_{\alpha \beta}$$$$\quad\Longleftrightarrow\quad \{X^A,P_B\}=-\{P_B,X^A\}=\delta^A_B\,,\quad \{\theta_i^A,\pi^j_B\}=\{\pi^j_B,\theta_i^A\}=\delta^A_B\delta^i_j\,.$$
The phase space coordinates $Z^A_\alpha$ are natural coordinates on the tensor
product ${\mathbb R}^{n,2}\otimes {\mathbb R}^{2|2s}$.
The algebra $\mathfrak{o}(n,2)$ is linearly realized on ${\mathbb R}^{n,2}\,\oplus\,\Pi({\mathbb R}^{n,2}\otimes{\mathbb R}^s)$ as
$$\textsc{J}^{AB}=\textsc{X}^A\textsc{P}^B-\textsc{X}^{B}\textsc{P}^A \,-\,i\,\vartheta^A_i\frac{\partial}{\partial \vartheta_B^i}\,+\,i\,\vartheta^B_i\frac{\partial}{\partial \vartheta_A^i}$$
The Weyl symbols of these generators of the	algebra $\mathfrak{o}(n,2)$
are the bilinears $$J^{AB}={\cal J}^{\alpha\beta}Z_\alpha^A Z^B_\beta$$
The 
Lie superalgebra $$\mathfrak{osp}(2s|2)=\mbox{span}
\{\textsc{t}_{\alpha\beta}\}$$ can be presented
\begin{itemize}
	\item by its {generators} $\textsc{t}_{\alpha\beta}=\textsc{t}_{\beta\alpha}$ (where $\alpha,\beta=1,2$)
	\item modulo the {graded commutation relations}
$$\left[\textsc{t}_{\alpha\beta},\textsc{t}_{\gamma\delta}\right]\,=\,i\,{\cal J}_{\beta\gamma}\textsc{t}_{\alpha\delta}\, +\, \mbox{(anti)symmetrizations}\,.$$
\end{itemize}
The Weyl symbols of the operators $\square\,$, $X\cdot\partial_X+\frac{n+2}2\,$, $X^2$, $\frac{\partial}{\partial X}\cdot \vartheta_i$,
 $\frac{\partial}{\partial X}\cdot \frac{\partial}{\partial \vartheta_i}$,
 $X\cdot \frac{\partial}{\partial \vartheta_i}$,
 $\frac{\partial}{\partial X}\cdot \frac{\partial}{\partial \vartheta_i}$.
 $\vartheta_i\cdot \frac{\partial}{\partial \vartheta_j}\,-\,\delta_i^j\,\frac{n+2}{2}$, $\vartheta_i\cdot \vartheta_j$, and  $\frac{\partial}{\partial \vartheta_i}\cdot \frac{\partial}{\partial \vartheta_j}$ are the bilinears
$$T_{\alpha\beta}=\eta_{AB}Z_\alpha^A Z^B_\beta$$
The Moyal graded commutators (or graded Poisson brackets) of generators reproduce the corresponding graded commutation relations.
The respective realizations of $\mathfrak{o}(n,2)$ and $\mathfrak{osp}(2s|2)$ are maximal commutants in the algebra of quadratic Weyl symbols:
they form a Howe dual pair.

\vspace{2mm}\noindent\textbf{Ambient construction of tensorial singletons \cite{Bekaert:2009fg}:} 
\textit{The spin $s\in\mathbb N$ singleton module ${\cal D}(s+\frac{n}2-1\,;\,s,\ldots,s)$ can be realized as a space of distributions on the superspace ${\mathbb R}^{n,2}\,\oplus\,\Pi({\mathbb R}^{n,2}\otimes{\mathbb R}^s)$ which are annihilated by the $\mathfrak{osp}(2s|2)$ superalgebra, which is Howe dual to $\mathfrak{o}(n,2)$ in the superalgebra of linear operators on ${\mathbb R}^{n,2}\,\oplus\,\Pi({\mathbb R}^{n,2}\otimes{\mathbb R}^s)$.}

\section{Higher-spin algebras:\\ various definitions}

\subsection{Higher-spin algebras as realizations\\ of universal enveloping algebras}

The simplest higher-spin algebra is the infinite-dimensional extension of the algebra $\mathfrak{o}(n,2)$ used by Vasiliev in his construction of a higher-spin gravity theory with $AdS_{n+1}$ as maximally symmetric solution \cite{Vasiliev:2003ev}, conjectured to be the holographic dual of free scalar singletons on $\partial AdS_{n+1}$ \cite{Konstein:2000bi}.

\vspace{2mm}
\noindent\textbf{Definition (Vasiliev, 2003):} \textit{The quotient of the universal enveloping algebra ${\cal U}\big(\mathfrak{o}(n,2)\big)$ by its annihilator on the scalar singleton module ${\cal D}(\frac{n}2-1;0)$ is the \textbf{$\bf AdS_{n+1}/CFT_n$ higher-spin algebra}.}\footnote{Strictly speaking, the higher-spin algebra defined in \cite{Vasiliev:2003ev} is a real form of the complex algebra considered here for the sake of simplicity (see e.g. the section 5 of the review \cite{Bekaert:2005vh} for more details).}

\vspace{2mm}
\noindent In other words, the complex $AdS_{n+1}/CFT_n$ higher-spin algebra is the realisation of the associative algebra ${\cal U}\big(\mathfrak{o}(n+2,{\mathbb C})\big)$ on the module ${\cal D}(\frac{n}2-1;0)$.\footnote{Due to the realization of $n=3$ (super)singletons in terms of oscillators (\textit{c.f.} footnote \ref{osc}), this $AdS_4/CFT_3$ higher-spin algebra can be defined more simply as a Weyl algebra.
The higher-spin (super)algebras were originally defined as the natural generalizations of this construction \cite{hsalg} (see also \cite{Gunaydin}).}

\subsection{Higher-spin algebras as centralizers of Howe dual algebras}

The previous definition, based on the scalar singleton, is equivalent to a purely algebraic definition, based on the above-mentioned $\mathfrak{sp}(2)$
subalgebra of $A_{n+2}$ (see e.g. \cite{Bekaert:2008sa} for a review of the proof).

\vspace{2mm}
\noindent\textbf{Theorem (Vasiliev, 2003):}\textit{
The centraliser ${\cal C}_{A_{n+2}}(\,\mathfrak{sp}(2)\,)$ of $\mathfrak{sp}(2)\subset A_{n+2}$ possesses two ideals spanned by the elements that also belong either to $\mathfrak{sp}(2)A_{n+2}$ or to $A_{n+2}\mathfrak{sp}(2)\,$.
The quotient of the centraliser by any of these ideals
is the $AdS_{n+1}/CFT_n$ {higher-spin algebra}.
}

\subsection{Higher-spin algebras as invariants of Howe dual algebras}

In order to translate the previous abstract definitions into a very concrete and explicit realization,
Vasilev made use \cite{Vasiliev:2003ev} of two major contributions of Weyl: symmetric symbol calculus and classical invariant theory (see e.g. \cite{Bekaert:2008sa} for more details).

\vspace{2mm}
\noindent\textbf{Theorem (Vasiliev, 2003):}\textit{
The $AdS_{n+1}/CFT_n$ {higher-spin algebra} is isomorphic to the algebra of $\mathfrak{sp}(2)$-invariant Weyl symbols $\in A_{n+2}$ modulo $\mathfrak{o}(n,2)$-traces. This space is spanned by the equivalence class of polynomials that depend on the phase space variables $Y^A_\alpha$ only through the combination $L^{AB}=\varepsilon^{\alpha\beta}Y_\alpha^A Y^B_\beta$, modulo the polynomials which are proportional to $U_{\alpha\beta}=\eta_{AB}Y_\alpha^A Y^B_\beta$. The restriction of the adjoint representation of the $AdS_{n+1}/CFT_n$ {higher-spin algebra}
to the $\mathfrak{o}(n,2)$ subalgebra is decomposable as the multiplicity free direct sum of all finite-dimensional  $\mathfrak{o}(n,2)$-modules labeled
by rectangular Young diagrams made of two rows.
}

\subsection{Higher-spin algebras as algebras\\ of symmetries}

The scalar singleton module ${\cal D}(\frac{n}2-1;0)$ may be realized as a space of distributions on ${\mathbb R}^{n,2}$ in the kernel of
the representation of the algebra $\mathfrak{sp}(2)\subset A_{n+2}$,
so a natural definition for its symmetries arises:
\begin{itemize}
	\item A \textbf{symmetry generator of the scalar singleton} is an element of the centralizer of $\mathfrak{sp}(2)$ in the space $A_{n+2}$ quotiented by the subspace $A_{n+2}\mathfrak{sp}(2)$, i.e. a differential operator $\textsc{o}$ on ${\mathbb R}^{n,2}$ that weakly commutes with all operators $\textsc{u}_{\alpha\beta}$ in the sense that: $$[\textsc{o}\stackrel{*}{,}\textsc{u}_{\alpha\beta}]=\textsc{P}_{\alpha\beta}^{\gamma\delta}\textsc{u}_{\gamma\delta}$$
with $\textsc{P}_{\alpha\beta}^{\gamma\delta}$ some coefficients.
	\item A \textbf{trivial symmetry generator of the scalar singleton} is a  symmetry generator that belongs to the left ideal $A_{n+2}\mathfrak{sp}(2)$, i.e. that vanishes on the kernel of the operators $\textsc{u}_{\alpha\beta}$.
	\item An \textbf{algebra of symmetries} is an algebra of equivalence classes of symmetry generators.  
\end{itemize}
The equation $\textsc{u}_{\alpha\beta}\Psi(X)=0$ for the distribution $\Psi$ on ${\mathbb R}^{n,2}$ describing the scalar singleton is preserved by the transformation $\Psi\mapsto \textsc{O}\Psi$ when $\textsc{O}$ is a symmetry generator, and $\textsc{O}\Psi=0$ when the symmetry generator is trivial.
The realisation of the universal enveloping algebra ${\cal U}\big(\mathfrak{o}(n,2)\big)$ on the scalar singleton module ${\cal D}(\frac{n}2-1;0)$ is automatically an algebra of such symmetry generators.
Indeed, one of the leitmotiv behind higher-spin symmetries is that, while super symmetries correspond to ``square roots'' of isometry generators, higher symmetries are ``powers'' of isometry generators.

\vspace{2mm}
\noindent\textbf{Theorem  (Eastwood, 2002):}\textit{
The $AdS_{n+1}/CFT_n$ {higher-spin algebra} is the subalgebra of the Weyl algebra $A_{n+2}$ of differential operators on the ambient space ${\mathbb R}^{n,2}$, spanned by (the equivalence classes of) the symmetry generators of the free scalar singleton.
}

\vspace{2mm}Let us finally turn back to the tensorial singletons.
The spin $s\in\mathbb N$ singleton module ${\cal D}(s+\frac{n}2-1\,;\,s,\ldots,s)$ may be realized as a space of distributions on the superspace
${\mathbb R}^{n,2}\,\oplus\,\Pi({\mathbb R}^{n,2}\otimes{\mathbb R}^s)$ annihilated by the $\mathfrak{osp}(2s|2)$ superalgebra, which
is Howe dual to $\mathfrak{o}(n,2)$.
Following the previous track, two natural definitions arise:
\begin{itemize}
	\item A \textbf{symmetry generator of a free spin $s$ singleton} is an element of the centralizer of $\mathfrak{osp}(2s|2)\subset A_{n+2|s(n+2)}$ in the
space $A_{n+2|s(n+2)}$ quotiented by the subspace $A_{n+2|s(n+2)}\mathfrak{osp}(2s|2)$, i.e. a differential operator that commutes
with all operators $\textsc{t}_{\alpha\beta}$ on their kernel.
	\item A \textbf{trivial symmetry generator of a spin $s$ singleton} is a symmetry generator in the subspace $A_{n+2|s(n+2)}\mathfrak{osp}(2s|2)$, so it vanishes on the singleton module.
\end{itemize}

With the help of two ingredients (the Howe duality and the Sergeev theory of classical Lie {super}algebra invariants), one can determine the

\vspace{2mm}\noindent\textbf{Singleton maximal symmetry algebra \cite{Bekaert:2009fg}:}\textit{
The following algebras are isomorphic:
\begin{itemize}
	\item The maximal algebra of symmetry generators of a free spin $s$ singleton
	\item The realization of the universal enveloping algebra ${\cal U}\big(\mathfrak{o}(n+2,{\mathbb C})\big)$ on the spin $s$ singleton module ${\cal D}(s+\frac{n}2-1\,;\,s,\ldots,s)$.
	\item The subalgebra of differential operators on the superspace ${\mathbb R}^{n+2|s(n+2)}$ whose Weyl symbols are $\mathfrak{osp}(2s|2)$-invariants modulo $\mathfrak{o}(n+2)$-traces.
\end{itemize}
Moreover, the adjoint representation of these algebras is a completely reducible $\mathfrak{o}(n,2)$-module which decomposes as the sum of all
irreducible $\mathfrak{o}(n,2)$-modules labeled by Young diagrams such that 
\begin{itemize}
	\item (i) all columns are of even length,
	\item (ii) the sum of the lengths of any two columns is not greater than $n+2$, and
	\item (iii) any column on the right of the $2s$-th column is
of length two,
\end{itemize}
where each such irreducible module appears with multiplicity one.
}

\vspace{2mm}\noindent
\textbf{Remarks}: These results
\begin{itemize}
	\item generalize the previous works on the particular cases: any $n\geqslant 3$ but $s=0$ \cite{Eastwood:2002su,Vasiliev:2003ev} or any $s\in{\mathbb N}/2$ but $n=4$ \cite{Vasiliev:2001zy}
	\item are formulated in terms of the above ambient definition of singleton symmetries, somewhat stronger than the local definitions adopted in \cite{Eastwood:2002su,Bekaert:2009fg}.\footnote{Indeed, for technical reasons a complete proof of the exhaustivity is lacking in \cite{Bekaert:2009fg} if one takes the local definition of singleton symmetries, but the proof is complete for the stronger definition.}
	\item can be summarized in a slogan (which will be my conclusion): 
	
	\textit{All symmetry generators of a free singleton are polynomials in the isometry generators.}
\end{itemize}

\section*{Acknowledgments}

I am grateful to I. Bandos, G. Barnich, N. Boulanger, M. Grigoriev, C. Iazeolla, M.~Rausch de~Traubenberg, D. Sorokin, P. Sundell, M. Tsulaia and M. Valenzuela for various enjoyable discussions and collaborations, scattered over the years, involving singletons and/or higher-spin algebras. E. Angelopoulos and M. Laoues are also thanked for useful exchanges. I am particularly grateful to C. Iazeolla for his patient reading of the manuscript and his useful comments. I also thank A. Raj for pointing out some typos.

I also acknowledge the organisers of the meetings mentioned in the front page for
their kind invitation to present talks on related results, with a special thank to the organizers of the ``7th spring school and workshop on quantum field theory \& Hamiltonian systems'' and of the ``6th mathematical physics meeting: summer school and conference on modern mathematical physics'' for giving me the opportunity to publish the present material (split into two distinct pieces) in their two respective proceedings \cite{BekaertCalimanesti}.



\clearpage

\begin{thebibliography}{B-B} 
\medskip
\begin{footnotesize} 

\bibitem{Dirac:1963ta}
P.~A.~M. Dirac, ``{A remarkable representation of the 3+2 de Sitter group},''
\textit{J. Math. Phys.} {\bf 4} (1963)  901.

\bibitem{DiFrancesco:1997nk}
P.~Di Francesco, P.~Mathieu and D.~Senechal,
\textit{Conformal Field Theory} (Springer, 1997) Section 4.1.
  
\bibitem{Dirac:1936fq}
P.~A.~M.~Dirac, ``Wave equations in conformal space,''
Annals Math. {\bf 37} (1936) 429.
	
\bibitem{Fefferman}
C.~Fefferman and C.~R.~Graham, ``Conformal invariants'' in \textit{\'Elie Cartan et les Math\'ematiques
d'Aujourdhui} (Ast\'erique, 1985) 95.

\bibitem{Siegel}
I.~T.~Todorov, M.~C.~Mintchev, V.~B.~Petkova, \textit{Conformal Invariance in Quantum Field Theory} (Scuola Normale Superiore di Pisa, 1978)  
Section I.3.B;\\
W.~Siegel, ``\textit{Fields},'' {\tt arXiv:hep-th/9912205}, Section 1.I.A.6.

\bibitem{Bekaert:2006py}
X.~Bekaert and N.~Boulanger, ``{The unitary representations of the Poincar\'e
group in any spacetime dimension},''
in the proceedings of the `Deuxi\`emes rencontres de
physique math\'ematique \`a Modave' (Modave, Belgium; August 2006)
{{\tt arXiv:hep-th/0611263}}.

\bibitem{Fuchs}
J.~Fuchs and C.~Schweigert, \textit{Symmetries, Lie Algebras and
Representations} (Cambridge University Press, 1997) Chapter
14.

\bibitem{Metsaev:1997nj}
R.~R.~Metsaev,
``Arbitrary spin massless bosonic fields in d-dimensional anti-de Sitter
space,''
Lect.\ Notes Phys.\  {\bf 524} (1997) 331
[{\tt arXiv:hep-th/9810231}].

\bibitem{deWit:1999ui}
B.~de Wit and I.~Herger, ``Anti de Sitter supersymmetry,'' Lect.\ Notes Phys.\  {\bf 541} (2000) 79
[{\tt arXiv:hep-th/9908005}] Section 5.

\bibitem{Iazeolla:2008ix}
F.~A.~Dolan,
``Character formulae and partition functions in higher dimensional  conformal
field theory,''
J.\ Math.\ Phys.\ {\bf 47} (2006) 062303
[{\tt arXiv:hep-th/0508031}] Section 2 and Appendix C;\\
C.~Iazeolla and P.~Sundell, ``{A fiber approach to harmonic analysis of
unfolded higher-spin field equations},''
JHEP {\bf 10} (2008)  022 [{\tt arXiv:0806.1942} [{\tt hep-th}] Appendices A.1 and A.2.

\bibitem{Flato:1978qz}
M.~Flato and C.~Fr\o nsdal, ``{One massless particle equals two Dirac singletons},''
{Lett. Math. Phys.} {\bf 2} (1978)  421.

\bibitem{Angelopoulos:1997ij}
E.~Angelopoulos and M.~Laoues, ``{Masslessness in $n$-dimensions},''
{Rev. Math. Phys.} {\bf 10} (1998)  271 [{\tt arXiv:hep-th/9806100}].

\bibitem{Gunaydin}
M.~Gunaydin, ``Singletons and doubleton supermultiplets of space-time groups and infinite spin
superalgebras,'' in \textit{Supermembranes and Physics in 2 + 1 Dimensions: Proceedings of the Trieste Conference (17-21 July 1989, ICTP, Trieste)} (World Scientific, 1990).

\bibitem{Engquist:2007vj}
J.~Engquist, P.~Sundell and L.~Tamassia,
``Singleton strings,''
Fortsch.\ Phys.\  {\bf 55} (2007) 711
[{\tt arXiv:hep-th/0701081}].

\bibitem{Ehrman}
J.B. Ehrman, Proc.\ Cambridge Phil.\ Soc.\ {\bf 53} (1957) 290.

\bibitem{Angelopoulos:1999bz}
E.~Angelopoulos and M.~Laoues,
``Singletons on AdS(n),''
Math.\ Phys.\ Studies\ {\bf 22} (2000) 3.

\bibitem{Majorana}
E.~Majorana,
``Teoria relativistica di particelle con momento
intrinseco arbitrario,''
Nuovo Cim.\  {\bf 9} (1932) 335 [in Italian].

\bibitem{Casalbuoni:2006fa}
R.~Casalbuoni,
``Majorana and the infinite component wave equations,''
PoS E {\bf MC2006} (2006) 004
[{\tt arXiv:hep-th/0610252}].

\bibitem{Bekaert:2009pt}
X.~Bekaert, M.~Rausch de~Traubenberg, and M.~Valenzuela, ``{An infinite supermultiplet of massive higher-spin fields},''
JHEP {\bf 05} (2009)  118 [{\tt arXiv:0904.2533 [hep-th]}].

\bibitem{Angelopoulos:1980wg}
E.~Angelopoulos, M.~Flato, C.~Fr\o nsdal and D.~Sternheimer,
``Massless particles, conformal group and de Sitter universe,''
Phys.\ Rev.\  D {\bf 23} (1981) 1278.

\bibitem{Siegel:1988gd}
W.~Siegel, ``{All free conformal representations in all dimensions},'' Int.\ J.\ Mod.\ Phys.\
A {\bf 4} (1989)  2015.

\bibitem{Metsaev:1995jp}
R.~R.~Metsaev,
``All conformal invariant representations of d-dimensional anti-de-Sitter
group,''
Mod.\ Phys.\ Lett.\  A {\bf 10} (1995) 1719.

\bibitem{Bandos:2005mb}
I.~Bandos, X.~Bekaert, J.~A. de~Azcarraga, D.~Sorokin, and M.~Tsulaia,
``{Dynamics of higher spin fields and tensorial space},'' JHEP\ {\bf 05}
(2005)  031 [{\tt arXiv:hep-th/0501113}] Section 2.

\bibitem{Gross:1964}
L.~Gross, ``Norm invariance of mass zero equations under the conformal group,'' 
J.\ Math.\ Phys.\ {\bf 5} (1964) 687.

\bibitem{Deser:1997mz}
S.~Deser, A.~Gomberoff, M.~Henneaux and C.~Teitelboim,
``Duality, self-duality, sources and charge quantization in abelian N-form
theories,''
Phys.\ Lett.\  B {\bf 400} (1997) 80
[{\tt arXiv:hep-th/9702184}].

\bibitem{Eastwood:2002su}
M.~G. Eastwood, ``{Higher symmetries of the Laplacian},'' Annals\ Math.\ {\bf 161} (2005) 1645 [{\tt arXiv:hep-th/0206233}].

\bibitem{Bars:2001xv}
I.~Bars,
``2T physics 2001,''
AIP Conf.\ Proc.\  {\bf 589} (2001) 18;
AIP Conf.\ Proc.\  {\bf 607} (2001) 17
[{\tt arXiv:hep-th/0106021}].

\bibitem{Bekaert:2009fg}
X.~Bekaert and M.~Grigoriev,
``Manifestly conformal descriptions and higher symmetries of bosonic singletons,''
SIGMA {\bf 6} (2010) 038
[{\tt arXiv:0907.3195} [{\tt hep-th}]].

\bibitem{Barnich:2004cr}
G.~Barnich, M.~Grigoriev, A.~Semikhatov and I.~Tipunin, ``Parent field theory
and unfolding in {BRST} first-quantized terms,'' Commun. Math. Phys.
{\bf 260} (2005)  147 [{\tt hep-th/0406192}].

\bibitem{tractor}
	T.~N.~Bailey, M.~G.~Eastwood and A.~R.~Gover, ``Thomas's structure bundle for conformal, projective and 
	related structures,'' Rocky Mountain J. Math. {\bf 24} (1994) 1191.

\bibitem{Gover:2009vc}
  A.~R.~Gover and A.~Waldron,
  ``The $\mathfrak{so}(d+2,2)$ minimal representation and ambient tractors: the conformal geometry of momentum space,''
  {\tt arXiv:0903.1394} [{\tt hep-th}];\\
  M.~Grigoriev and A.~Waldron,
  ``Massive Higher Spins from BRST and Tractors,''
  Nucl.\ Phys.\  B {\bf 853} (2011) 291
  [{\tt arXiv:1104.4994} [{\tt hep-th}]].

\bibitem{Arvidsson:2006fq}
P.~Arvidsson and R.~Marnelius, ``{Conformal theories including conformal
  gravity as gauge theories on the hypercone},''
	{{\tt arXiv:hep-th/0612060}}, Sections 5 and 6.

\bibitem{Bastianelli:2008nm}
F.~Bastianelli, O.~Corradini and E.~Latini,
``Spinning particles and higher spin fields on (A)dS backgrounds,''
JHEP {\bf 0811} (2008) 054
[{\tt arXiv:0810.0188} [{\tt hep-th}]].

\bibitem{Bekaert:2005vh}
X. Bekaert, S. Cnockaert, C. Iazeolla and M.A.
Vasiliev, ``Nonlinear higher spin theories in various
dimensions,'' in the proceedings of the `First Solvay Workshop on
Higher-Spin Gauge Theories' (Brussels, Belgium; May 2004) [{\tt
arXiv:hep-th/0503128}].

\bibitem{Vasiliev:2003ev}
M.~A. Vasiliev, ``{N}onlinear equations for symmetric massless higher spin
fields in ({A})d{S}(d),'' {\em Phys. Lett.} {\bf B567} (2003)  139
[{\tt hep-th/0304049}].

\bibitem{Konstein:2000bi}
S.~E.~Konstein, M.~A.~Vasiliev and V.~N.~Zaikin,
``Conformal higher spin currents in any dimension and AdS/CFT
correspondence,''
JHEP {\bf 0012} (2000) 018
[{\tt arXiv:hep-th/0010239}];\\
E.~Sezgin and P.~Sundell,
``Massless higher spins and holography,''
Nucl.\ Phys.\  B {\bf 644} (2002) 303
[{\tt arXiv:hep-th/0205131}];\\
I.~R.~Klebanov and A.~M.~Polyakov,
``AdS dual of the critical O(N) vector model,''
Phys.\ Lett.\  B {\bf 550} (2002) 213
[{\tt arXiv:hep-th/0210114}].

\bibitem{hsalg}
S.~E.~Konstein, M.~A.~Vasiliev~, ``Massless representations and admissibility condition for higher spin superalgebras,''
Nucl.\ Phys.\ B {\bf 312} (1989) 402.
  
\bibitem{Bekaert:2008sa}
X.~Bekaert,
``Comments on higher-spin symmetries,''
Int.\ J.\ Geom.\ Meth.\ Mod.\ Phys.\  {\bf 6} (2009) 285
[{\tt arXiv:0807.4223} [{\tt hep-th}]] Section 3.

\bibitem{Vasiliev:2001zy}
M.~A. Vasiliev, ``Conformal higher spin symmetries of {4D} massless
supermultiplets and $\mathfrak{osp}(L,2M)$ invariant equations in generalized
(super)space,'' Phys.\ Rev.\ {\bf D66} (2002)  066006 [{\tt hep-th/0106149}];\\
S.~C. Anco and J.~Pohjanpelto, ``{Symmetries and currents of massless neutrino fields, electromagnetic and graviton fields},'' CRM\ Proc.\ and Lect.\
Notes\ {\bf 34} (2004)  1
[{\tt arXiv:math-ph/0306072}];
``{Generalized symmetries of massless free fields on Minkowski space},'' {\em SIGMA} {\bf 4} (2008) 004[{\tt arXiv:0801.1892} [{\tt math-ph}].

\bibitem{BekaertCalimanesti}
X.~Bekaert, ``The many faces of singletons,'' Physics\ Ann.\ Univ.\ Craiova\ {\bf 21} 
(2011) 154; ``Singletons and their maximal symmetry algebras'' to appear in \textit{Proceedings of the `6th mathematical physics meeting: summer school and conference on modern mathematical physics'}.

\end{footnotesize} 
\end{thebibliography}
\end{document}